\documentclass[aps,prb,twocolumn,superscriptaddress,amsmath,amssymb,floatfix]{revtex4-1}

\usepackage{graphics}
\usepackage{epsfig}
\usepackage{bm}
\usepackage{color}

\begin{document}

\graphicspath{{Figures/}}

\title{Hall conductivity in the normal and superconducting phases of the Rashba system with Zeeman field}
\author{Suk Bum Chung}
\affiliation{Center for Correlated Electron Systems, Institute for Basic Science (IBS), Seoul 151-747, Korea}
\affiliation{Department of Physics and Astronomy, Seoul National University, Seoul 151-747, Korea}
\affiliation{Department of Physics and Astronomy, University of California Los Angeles, Los Angeles, California 90095-1547, USA}
\author{Rahul Roy$^3$}

\date{\today}

\begin{abstract}
We study the intrinsic Hall conductivity of the ordinary and topological  superconducting phases of a Rashba metal  in a perpendicular Zeeman field. In this system the normal metal breaks time reversal symmetry while the superconducting order parameter does not, in contrast to the chiral p-wave superconducting state predicted in the monolayer strontium ruthenate (Sr$_2$RuO$_4$) whose Hall conductivity has been studied extensively.  We study the effects of  intra-band and inter-band pairing and find there is qualitatively larger change in the intrinsic Hall conductivity when there is inter-band pairing, with the change in magnitude linear in the pairing gap. We argue that inter-band pairing leads in general to higher energy costs for the topological phase compared to the topologically trivial phase and thus that the qualitative behavior of the intrinsic Hall conductivity with superconductivity in these systems could provide  important clues about the nature of pairing in the superconducting phase and even some hints of whether it is topological or not.  
\end{abstract}

\maketitle

\section{Introduction}

Although the recent interest in intrinsic Hall conductivity largely focuses on its remarkable quantization in insulators \cite{Laughlin1981,Thouless1982,Haldane1988,Liu2008,Chang2013}, a nonzero Hall conductivity is possible for any systems, including metal and superconductor, that breaks time-reversal symmetry. While there is no quantization of the Hall conductivity in metals, there nonetheless exists the identical geometric picture of it through the Karplus-Luttinger formula \cite{Karplus1954,Haldane2004}, which states that the intrinsic Hall conductivity is proportional to the net Berry curvature in the Brillouin zone. 
The (non-)quantization of the Hall conductivity in insulators (metals) can be explained by this formula together with the fact that the total Berry curvature for each band in the first Brillouin zone is quantized. On the other hand, less has been known about what determines the magnitude of intrinsic Hall conductivity of superconductors, in spite of recent detection of time-reversal symmetry breaking in various unconventional superconductors including not only Sr$_2$RuO$_4$ \cite{Xia2006} but also more recently UPt$_3$ \cite{Schemm2014} and URu$_2$Si$_2$ \cite{Schemm2013, Schemm2014}.

 The possibility of exotic physics in topological superconductors such as non-abelian statistics of vortex defects  has led to a great deal of theoretical and experimental interest in possible candidate materials. Sr$_2$RuO$_4$ \cite{Mackenzie2003} which is thought to be a chiral $p_{x}+ i p_{y}$ superconductor has perhaps attracted the most interest. Evidences for broken time-reversal symmetry in Sr$_2$RuO$_4$  include muon spin relaxation measurements \cite{Luke1998} in addition to the Kerr rotation. 
  
The Kerr rotation angle is a measure of the intrinsic Hall conductivity and a non-zero value of this angle indicates time-reversal symmetry breaking. Somewhat surprisingly however, theoretical calculations show that in the long wavelength limit, the intrinsic,  {\it i.e.} impurity-independent, Hall conductance of a pure $p+ ip$ superconductor is zero~\cite{Lutchyn2008,Roy2008}. A non-zero intrinsic Hall conductivity has only been obtained so far in chiral $p$-wave models that allow for interband pairing \cite{Edward2012,Gradhand2013,Taylor2013}. Unlike in insulators, where the Hall conductance is quantized, the Hall conductance of superconductors

A different type of time-reversal symmetry breaking superconductor that is attracting widespread interest in recent years is the the one  that occurs in 2D metal with strong Rashba spin-orbit coupling under a perpendicular Zeeman field. In this system, a topologically non-trivial superconducting phase analogous to the spin-polarized chiral $p$-wave superconductor can arise for the right range  of the chemical potential and the Zeeman field 
\cite{Sau2010,Alicea2010,Oreg2010,Mourik2012,Das2012}.  There are two key differences between this Rashba superconductor and the chiral $p$-wave superconductor that affect the Kerr rotation angle. The first is that the time-reversal symmetry breaking in the Rashba superconductor does not originate from the Cooper pairing and is already present in the normal phase. The other is that the Rashba system is inherently multi-band due to the spin-orbit coupling splitting of the Fermi surface, naturally raising the question whether the Cooper pairing is purely intraband or has a nonzero interband component. The possibility of the inter-band pairing has been discussed in the recent literature \cite{Sato2009,Alicea2010} and it been  noted that its presence or absence will not affect the possibility of the topological quantum phase transition.  However, the physical consequence of the inter-band pairing has remained an under-investigated aspect of the field.

In this paper, we calculate the Hall conductivity of the Rashba superconductor in which an interesting interplay of time-reversal symmetry breaking in the normal phase and a time-reversal invariant order parameter occurs. We separately consider the effects of the intra-band and inter-band pairing and  find that the effect of superconductivity on the Hall conductivity is qualitatively stronger when there is nonzero inter-band pairing. A more precise statement of our result is that the change in the Hall conductivity due to superconductivity is linear in the pairing gap with a nonzero inter-band pairing but quadratic in pairing gap with a purely intra-band pairing. The effect of inter-band pairing is consistent with the recent calculations of the Hall conductivity in multi-band chiral $p$-wave superconductor models. 

This paper is organized as follows. In the section II, we calculate the Hall conductivity of the Rashba metal under a Zeeman field using the linear response and show that the result agrees with the Karplus-Luttinger formula. In the sections III-IV, we calculate the effect of Cooper pairing on the Hall conductivity of this system with purely intra-band pairing and with both intra- and inter-band pairing, respectively. In the section V, we discuss how the two cases would correspond to the topology of the superconducting phase followed by  a discussion in the conclusion.

\section{The 2D Rashba metal and its Hall conductivity}

 We use the term ``2d Rashba metal'' to describe a two dimensional system with a strong Rasha spin orbit coupling governed by an effective Hamiltonian of the form:
\begin{equation}
\hat{\mathcal{H}} =\frac{p^{2}}{2m^{*}}  - \mu - \alpha(p_{y} \sigma_1 +  p_{x}\sigma_2)
\end{equation}
where $m^{*}$ is the effective mass, $\alpha$ a  parameter which characterizes the strength of the spin-orbit coupling, and $\sigma_{1,2}$ are the Pauli spin matrices; for convenience, we will set $\hbar=1$. 

The two bands of the 2d Rashba metal have energies, $p^{2}/2m^{*}  - \mu \pm \alpha p $ and a Dirac like crossing at $\mathbf{p}=0$. 
The system is time-reversal invariant which implies that the Hall conductance is zero. This in turn means that the net or integrated Berry curvature over all $\mathbf{p}$ of all negative energy eigenstates for is zero. 
  
 In the presence of an effective Zeeman term (as could possibly be induced from a tunneling from a magnetic insulator), this picture changes. Time-reversal symmetry is no longer preserved, raising the possibility of 
a non-zero Hall conductance. 
We shall now confirm this possibility with a detailed calculation. The Hamiltonian with an effective Zeeman term $h\sigma_3$ can be written as 
\begin{equation}
\hat{\mathcal{H}} = \frac{p^{2}}{2m^{*}}  - \mu - {\bf d}_{\bf p}\cdot{\bm \sigma},
\end{equation}
where  ${\bf d} = (-\alpha p_y, \alpha p_x, h)$. The current operator for this Hamiltonian is
\begin{align}
\hat{v}_x =&  \frac{p_x}{m^*} -\sigma_2 \alpha,\nonumber\\
\hat{v}_y =&  \frac{p_y}{m^*} + \sigma_1\alpha.
\label{EQ:current}
\end{align}
and the finite temperature Green function is 
\begin{align}
\hat{\mathcal{G}}({\bf k},i\omega_n) &=[ i\omega_n - (\xi_k - {\bf d}_{\bf k}\cdot{\bm \sigma})]^{-1}\nonumber\\ 
&= \frac{\hat{P}_+(k)}{i\omega_n -(\xi_k-d_k)} + \frac{\hat{P}_-(k)}{i\omega_n -(\xi_k+d_k)},
\label{EQ:normalGreen}
\end{align}
where $\hat{P}_\pm(k) = (1  \pm{\bf d}_{\bf k}\cdot{\bm \sigma})/2$ are the band projection operators, $\omega_n$ is the Matsubara frequency, and $\xi_k \equiv k^2/2m^* - \mu$. We can then compute the optical Hall conductivity at $T=0$ using the Kubo formula:
\begin{align}
\sigma_{xy}(\omega) =& \frac{ie^2}{2\omega}\int \frac{d^2 k d\nu}{(2\pi)^3} {\rm tr}[\hat{v}_x \hat{\mathcal{G}}({\bf k},i\nu+\omega)\hat{v}_y \hat{\mathcal{G}}({\bf k},i\nu)]-(x \leftrightarrow y)\nonumber\\
=& -\frac{e^2 \alpha^2 h}{2\omega} \int \frac{d^2 k}{(2\pi)^2}\sum_{s=\pm}\frac{1}{d_k(2d_k+s\omega)}\nonumber\\&\times\int\frac{d\nu}{2\pi}\left[\frac{1}{i\nu -(\xi_k + sd_k)}-\frac{1}{i\nu+\omega -(\xi_k - sd_k)}\right],
\label{EQ:Kubo}
\end{align}
where we  have used
\begin{align}
& \int \frac{d^2 k d\nu}{(2\pi)^3} {\rm tr}[\hat{v}_x \hat{\mathcal{G}}({\bf k},i\nu+\omega)\hat{v}_y \hat{\mathcal{G}}({\bf k},i\nu)] - (x \leftrightarrow y)\nonumber\\
=& \int \frac{d^2 k d\nu}{(2\pi)^3} \sum_{s=\pm}\frac{\hat{v}_x \hat{P}_s\hat{v}_y \hat{P}_{-s}}{[i\nu +\omega -(\xi_k - sd_k)][i\nu -(\xi_k + sd_k)]}\nonumber\\ 
&- (i \leftrightarrow j)
\label{EQ:Kubo2}
\end{align}
and $k_{F\pm}$ are the momenta where $\xi_k \mp d_k =0$. This formula comes out simplest for $\omega=0$, giving us the Hall conductivity:
\begin{align}
\sigma_{xy}(\omega=0) =& e^2 \alpha^2 h \int \frac{d^2 k}{(2\pi)^2}\sum_{s=\pm}\frac{s}{4d_k^3} {\rm sgn}(\xi_k + s d_k)\nonumber\\ =& \frac{e^2 \alpha^2 h}{8\pi}\int^{k_{F+}}_{k_{F-}} \frac{2k dk}{(h^2 + \alpha^2 k^2)^{3/2}}\nonumber\\ =& \frac{e^2}{4\pi}h \left[\frac{1}{\sqrt{h^2 + \alpha^2 k^2_{F-}}}-\frac{1}{\sqrt{h^2 + \alpha^2 k^2_{F+}}}\right].
\label{EQ:HallNormal}
\end{align}
For the case where we only have a single Fermi surface, we can just set $k_{F-}=0$. Note that this requires $\mu \geq |h|$.

 This result could also have been obtained from the Karpus-Luttinger formula~\cite{Karplus1954,Haldane2004} for the Hall conductivity of a metal. The Karpus Luttinger formula states that
\begin{equation}
\sigma_{xy} = \frac{e^2}{2\pi}\int \frac{d^2 k}{(2\pi)^2} \sum_n \mathcal{F}^{xy}_n n_n ({\bf k})
\label{EQ:KL}
\end{equation}
where $n$ is the band index, $\mathcal{F}^{xy}_n$ is the Berry curvature of the $n$-th band, and $n_n ({\bf k})$ is the occupation number of the $n$-th band at momentum ${\bf k}$.

 It is straightforward to use this formula if one notes that the Berry curvature is entirely determined by the spin-dependent terms, 
hence is equal to the skyrmion density of the spin, ${\bf \hat{d}} \cdot (\partial_{k_x} {\bf \hat{d}} \times \partial_{k_y} {\bf \hat{d}})$. 
Thus, there is cancellation between the contribution from the larger Fermi pocket which covers the sphere starting from the north pole down to the `altitude' $\beta_+ \equiv \tan ^{-1} \alpha k_{F+}/h$ and that from the smaller Fermi pocket 
which starts from the south pole up to 
$\beta_- \equiv \tan ^{-1} \alpha k_{F-}/h$, 
giving us
\begin{align}
\sigma_{xy} =& \frac{e^2}{2\pi}\frac{1}{4\pi}\int d\phi \left(\int^1_{\cos \beta_+}- \int^{\cos \beta_-}_{-1}\right)d\cos\beta\nonumber\\
=&\frac{e^2}{4\pi}\left[\frac{h}{\sqrt{h^2 + \alpha^2 k^2_{F-}}}-\frac{h}{\sqrt{h^2 + \alpha^2 k^2_{F+}}}\right].
\end{align}
in agreement with the result obtained from the Kubo formula. 

\section{Pairing in the Rashba superconductor with Zeeman field}
 
Much of interest in the Rashba metal with an effective Zeeman arises from 
the existence of the topologically non-trivial superconducting phase. A simple form of pairing  
that has been studied extensively \cite{Fu2008,Sato2009,Sau2010,Alicea2010,Oreg2010} in this context 
is 
\begin{equation}
H_{pair} = |\Delta| \sum_{\bf k} c^\dagger_{{\bf k}\uparrow} c^\dagger_{-{\bf k}\downarrow} + {\rm h.c.}
\label{EQ:sWave} 
\end{equation}
The full first-quantized Hamiltonian including pairing is then
 \begin{equation}
\hat{\mathcal{H}}_{BdG} = \tau_3  ( \xi_k -  {\bm \sigma} \cdot {\bf d}_\parallel) -  \sigma_3 d_z  + \tau_1  |\Delta|
\end{equation}
in the Nambu basis $(\psi_{{\bf k},\uparrow}, \psi_{{\bf k},\downarrow}, -\psi^\dagger_{-{\bf k},\downarrow}, \psi^\dagger_{-{\bf k},\uparrow})^T$, with ${\bm \tau}$ being the Pauli matrices in the electron-hole space. This can be transformed to the band basis (with $\sigma_3$ diagonal with respect to the bands) as \cite{Sato2009,Alicea2010,Oreg2010}
\begin{align}
\hat{U}^\dagger \hat{\mathcal{H}}_{BdG} \hat{U} =& \tau_3  ( \xi_k - \sigma_3 d_k) - |\Delta| (-\tau_1  \hat{k}_y +  \tau_2  \sigma_3 \hat{k}_x)\sqrt{1-d_z^2}\nonumber\\ &+ |\Delta| \tau_2  \sigma_2  \hat{d}_z,
\label{EQ:sWaveBand}
\end{align}
where $\hat{U}$ gives us the basis transformation between the band basis and the original spin basis $(\psi_{{\bf k},\uparrow}, \psi_{{\bf k},\downarrow}, -\psi^\dagger_{-{\bf k},\downarrow}, \psi^\dagger_{-{\bf k},\uparrow})^T$. 
The pairing has both inter-band and intra-band components.  We would like to disentangle the contributions to the Hall conductivity from the two components and ask if there are qualitative differences between them. Interband pairing is of course  more likely to arise in systems where the superconductivity is intrinsic, {\it i.e.} 
not 
induced through the proximity effect. 
  
In the following two sections, we will calculate the Hall conductivity of the Rashba system in the superconducting state. We will show that the change in the Hall conductivity from its normal state value will 
depend qualitatively on absence or presence of significant interband pairing. 
The effect of superconductivity is weak in the absence of the interband pairing, in the sense that 
the Hall conductivity still retains some similarity to the Karplus-Luttinger formula of Eq.\eqref{EQ:KL}. 
 The purely intra-band Cooper pairing in the Rashba system is expected to break time-reversal symmetry, yet, strikingly, it does not significantly impact the Hall conductivity. This is in some sense consistent with the theoretical results for the superconducting phase of Sr$_2$RuO$_4$, where 
the intrinsic Hall conductivity in absence of the interband pairing is zero \cite{Lutchyn2008,Roy2008}, which is to say, the intrinsic Hall conductivity retains the normal state value in absence of the interband pairing. On the other hand, as we shall see later, there is significant  impact from time-reversal symmetry preserving interband pairing
. 

\subsection{The case of purely intra-band pairing}

We consider the purely intra-band pairing model that has an anti-chiral (chiral) pairing gap for the larger (smaller) Fermi surface,
\begin{equation}
\hat{\Delta}_\pm = |\Delta_\pm|(-\tau_1 \hat{k}_y \pm \tau_2 \hat{k}_x) \exp(\mp i\tau_3 \phi/2)
\label{EQ:intraGap}
\end{equation} 
($\phi$ being the phase difference between the two gaps), in the band basis, as this can be regarded as the purely intraband pairing that is closest to the $s$-wave pairing of Eq.\eqref{EQ:sWave} \cite{Fu2008,Sato2009,Sau2010,Alicea2010,Oreg2010}. In fact, this pairing with $|\Delta_+| = |\Delta_-|$ and $\phi=0$ is exactly equal to Eq.\eqref{EQ:sWaveBand} minus the interband pairing term $|\Delta| \tau_2  \sigma_2  \hat{d}_z$ 
 \footnote{Conversely, it has been shown \cite{Sato2009,Qi2010} that the $\phi = \pi$ gives us the time-reversal invariant topological superconductor in the $h \to 0$ limit.}, as can be shown from the full first quantized BdG Hamiltonian with the intraband pairing of Eq.\eqref{EQ:intraGap} 
 in the band basis: 
\begin{equation}
\hat{U}^\dagger \hat{\mathcal{H}}'_{BdG} \hat{U} = \tau_3 ( \xi_k - \sigma_3 d_k) + \hat{P}'_+  \hat{\Delta}_+ + \hat{P}'_-  \hat{\Delta}_-,
\end{equation}
where $\hat{P}'_\pm =  (1 \pm \sigma_3)/2$ are the band projection operators. The  operators $\hat{P}'_\pm =  (1 \pm \sigma_3)/2$ can be related to the normal state band projection operators $\hat{P}_\pm = (1 \pm {\bm \sigma}\cdot{\bf d})/2$ through
$$U^\dagger [ (1 \pm {\bm \sigma} \cdot \hat{\bf d}_\parallel) + \tau_3  \sigma_3 \hat{d}_z] U/2 = \hat{P}'_\pm.$$


Topologically, the purely intraband pairing gap of this subsection is equivalent to  the pure $s$-wave pairing, {\it i.e.} it gives us the same topological phases with the same Read-Green class of the topological quantum phase transition \cite{Read2000}. Since the inner and outer Fermi surfaces form two independent superconductors in the intraband pairing model, when there are two Fermi surfaces Eq.\eqref{EQ:intraGap} gives us a topologically trivial superconductivity, as we have two weak pairing superconductors whose topological invariants cancels out to zero due to their opposing chirality. When the inner Fermi surface vanishes to ${\bf k} = 0$ at $\mu = |h|$, $\hat{\Delta}_-$ vanishes as there is no degeneracy at ${\bf k} = 0$, and the topological phase transitions between the topologically trivial and non-trivial phases occur; note that the phase transition point for the purely $s$-wave pairing is slightly shifted to $\mu^2 = h^2 - |\Delta|^2$ 
\cite{Sau2010,Alicea2010,Oreg2010}. Given that the nonzero interband pairing was required for the purely $s$-wave pairing, it is natural that the purely intraband pairing of Eq.\eqref{EQ:intraGap} in general gives us a mixture of the $s$-wave and the $p$-wave pairing in the original spin basis \cite{Sato2009}:
\begin{widetext}
\begin{align}
\hat{U} (\hat{P}'_+  \hat{\Delta}_+ + \hat{P}'_- \hat{\Delta}_-) \hat{U}^\dagger =&-\bar{|\Delta|}\left[(\tau_1  \sqrt{1-\hat{d}_z^2} + \tau_2 {\bm \sigma} \cdot {\bf \hat{k}} \hat{d}_z)\cos\frac{\phi}{2} + \tau_2 ({\bm \sigma} \times {\bf \hat{k}}) \cdot {\bf \hat{z}} \sin\frac{\phi}{2}\right]\nonumber\\
&+\delta|\Delta|\left[\tau_1 ({\bm \sigma} \times {\bf \hat{k}}) \cdot {\bf \hat{z}} \cos\frac{\phi}{2} + (- \tau_1  {\bm \sigma} \cdot {\bf \hat{k}} \hat{d}_z + \tau_2  \sqrt{1-\hat{d}_z^2} ) \sin\frac{\phi}{2}\right]
\label{EQ:gapTransf}
\end{align}
where $\bar{|\Delta|} \equiv (|\Delta_+|+|\Delta_-|)/2, \delta|\Delta| \equiv (|\Delta_+|-|\Delta_-|)/2$ and the following transformation between the band basis and the spin basis is used:
\begin{align}
\hat{U}(-\tau_1  \hat{k}_y + \tau_2  \sigma_3 \hat{k}_x)\hat{U}^\dagger =& -\tau_1  \sqrt{1-\hat{d}_z^2} - \tau_2  {\bm \sigma} \cdot {\bf \hat{k}} \hat{d}_z,\nonumber\\
\hat{U}(-\tau_1 \hat{k}_y + \tau_2  \sigma_3 \hat{k}_x)\sigma_3 \hat{U}^\dagger =& \tau_1 ({\bm \sigma} \times {\bf \hat{k}}) \cdot {\bf \hat{z}}.
\label{EQ:intraTransf}
\end{align} 
\end{widetext}
Note that in Eq.\eqref{EQ:gapTransf}, the pairing gap breaks time-reversal symmetry due to 
the perpendicular Zeeman field, {\it i.e.} $\hat{d}_z = h/\sqrt{h^2+\alpha^2 k^2} \neq 0$ and the phase difference $\phi$ between the gaps of the two Fermi surfaces. 

In the case of the purely intraband pairing, the Hall conductivity calculation for the superconducting phase is no more complicated than the same calculation for the normal state. 
The Kubo formula provides the simplest gauge invariant method for calculating the optical Hall conductivity for a superconductor phase~\cite{Schrieffer1983,Lutchyn2008,Roy2008,Edward2012,Gradhand2013}.
When the pairing is purely intraband,  the Green function, 
the most important ingredient of the Kubo formula, 
remains block-diagonal in the band basis for the superconducting phase: 
\begin{align}
\hat{U}^\dagger \hat{G}_{BdG} \hat{U} \equiv& \hat{U}^\dagger[i\omega_n -  \hat{\mathcal{H}}'_{BdG}]^{-1} \hat{U}\nonumber\\
=  &-\frac{\hat{P}'_+ [  i\omega_n + \tau_3 (\xi_k-d_k)+\hat{\Delta}_+] }{\omega^2_n +E_+^2}\nonumber\\ 
&-\frac{\hat{P}'_- [  i\omega_n + \tau_3 (\xi_k+d_k)+\hat{\Delta}_-]}{\omega^2_n +E_-^2},
\label{EQ:BdG-Green}
\end{align}
where $E_\pm^2 = (\xi_k \mp d_k)^2 + |\Delta_\pm|^2$ is the quasi-particle eigenenergy. 
The Kubo formula for the superconducting phase can be obtained by inserting the BdG Green function of Eq.\eqref{EQ:BdG-Green} into Eq.\eqref{EQ:Kubo} with an overall factor of 1/2 to cancel out the BdG doubling:
\begin{align}
\sigma_{xy}(\omega) =& \frac{ie^2}{4\omega}\!\int \frac{d^2 k d\nu}{(2\pi)^3}\! {\rm tr}[\hat{v}_x \hat{G}_{BdG}({\bf k},i\nu+\omega)\hat{v}_y \hat{G}_{BdG}({\bf k},i\nu)]\nonumber\\
&-(x \leftrightarrow y)\nonumber
\end{align} 
at $T=0$. 
 Its important to note that the current operators have the same expression as in the normal state, {\it i.e.},  they do not get any contribution from the pairing terms of the Hamiltonian. 

\begin{figure}[h]
\includegraphics[width=\linewidth]{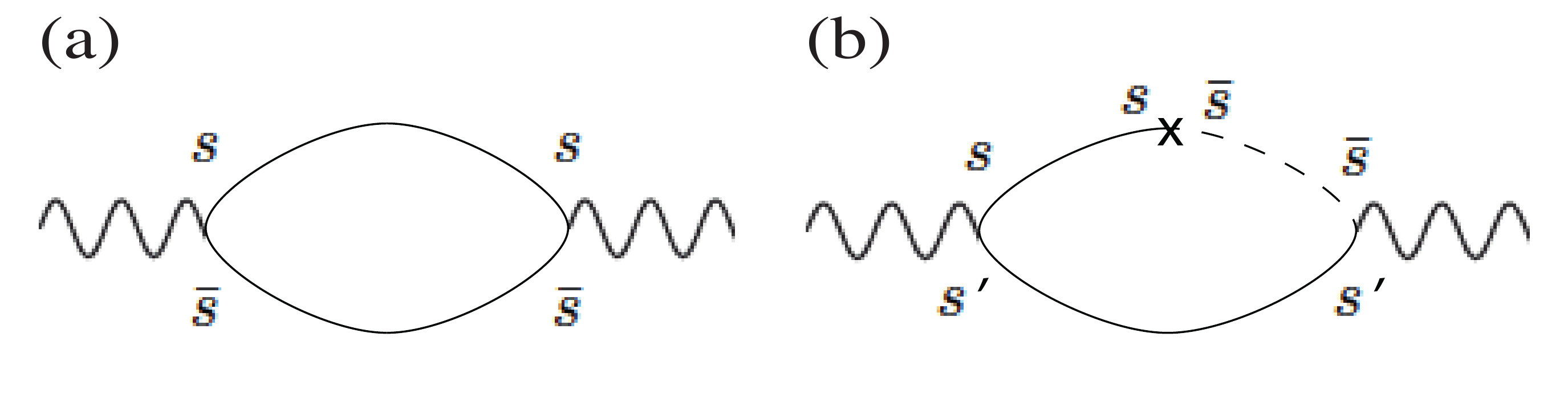}
\caption{Contribution to intrinsic Hall conductivity in the band basis, where $s,\bar{s},s'$ label bands. Filled and dotted curves denote the normal and anomalous Green function respectively. In both the normal state and the superconducting state with purely intraband pairing, there is contribution only from the diagram (a), which originate from the propagation of the electron in the upper band and the hole in the lower band (or vice versa for the superconducting state). However, the diagram (b) shows that with a nonzero interband pairing, possibility of interband propagation - this particular diagram involves the interband normal propagation coming from the combination of the intraband normal and anomalous propagation with interband pairing - gives rise to additional processes contributing to the Hall conductivity.}
\label{FIG:HallDiagram}
\end{figure}

We find that the effect of the Cooper pairing on the Hall conductivity can be attributed solely to the change in the quasiparticle spectrum in the sense that there is contribution only from the normal part of the BdG Green function $\hat{g} ({\bf k},i\omega) \equiv -\sum_{s=\pm}  \hat{U} [i\omega_n + \tau_3 (\xi_k - s d_k)] \hat{U}^\dagger/(\omega_n^2 + E_s^2)$ but none from the anomalous part 
$\hat{f}({\bf k},i\omega) \equiv -\sum_{s=\pm} \hat{U}\hat{P}'_s\hat{\Delta}_s\hat{U}^\dagger/(\omega_n^2 + E_s^2)$. In other words, the Hall conductivity still originates from the same process as in the normal state, the normal propagation of an electron and a hole from different bands as represented by (a) of Fig.\ref{FIG:HallDiagram}. Hence the time-reversal symmetry breaking of the intraband pairing playing no role, which includes lack of any dependence on the phase difference $\phi$ between the gaps of the two Fermi surfaces. \footnote{Hence we predict that the fluctuation of $\phi$ will not contribute to the Hall conductivity.} 
 The vanishing of the contribution from the anomalous part of the Green function is because
\begin{align}
& {\rm tr}\!\left[ \!\left(\frac{k_x}{m^*} -\alpha \sigma_2 \!\right)\! \hat{U}\hat{P}'_\pm\hat{\Delta}_\pm\hat{U}^\dagger \!\left(\frac{k_y}{m^*} + \alpha \sigma_1 \!\right)\!\hat{U}\hat{P}'_\mp\hat{\Delta}_\mp \hat{U}^\dagger \right]\nonumber\\
&- (x \leftrightarrow y)\nonumber\\
=& \pm  i\alpha^2 [\hat{k}_x^2 \hat{k}_y^2 (1-\hat{d}_z)^2 + (\hat{k}_x^2 \hat{d}_z + \hat{k}_y^2)(\hat{k}_x^2 + \hat{k}_y^2 \hat{d}_z)]\nonumber\\
&\times{\rm tr}[\tau_3 \hat{\Delta}_\pm \hat{\Delta}_\mp],   
\end{align}
where we first traced out band indices using 
\begin{align}
\hat{U}^\dagger \sigma_1 \hat{U} =&  - \tau_3  \sigma_1 (\hat{k}_x^2 + \hat{k}_y^2 \hat{d}_z) - \sigma_2 \hat{k}_x \hat{k}_y (1-\hat{d}_z)\nonumber\\
&-\sigma_3 \hat{k}_y \sqrt{1-\hat{d}_z^2},\nonumber\\
\hat{U}^\dagger \sigma_2 \hat{U} =&  - \tau_3 \sigma_1  \hat{k}_x \hat{k}_y (1-\hat{d}_z)  - \sigma_2 (\hat{k}_x^2 \hat{d}_z  + \hat{k}_y^2)\nonumber\\
&+ \sigma_3 \hat{k}_x \sqrt{1-\hat{d}_z^2}, 
\end{align}
vanishes upon angular integration as
\begin{equation}
{\rm tr}[\tau_3 \hat{\Delta}_\pm \hat{\Delta}_\mp]  = \pm i |\Delta_+| |\Delta_-| [2\hat{k}_x\hat{k}_y \cos \phi -(\hat{k}_x^2 - \hat{k}_y^2)\sin \phi]
\end{equation}
is odd with respect to the $\pi/4$ rotation. 
Therefore, for the superconducting state with the purely intraband pairing, we are left with the Hall conductivity of
\begin{align}
\sigma_{xy} 
=& \frac{e^2 \alpha^2 h} {2\pi} \int k dk \frac{1}{d_k} \frac{1}{(E_-+E_+)^2}\left( \frac{\xi_k + d_k}{E_-} -\frac{\xi_k - d_k}{E_+}\right).
\label{EQ:HallIntra}
\end{align}
Comparison between the integrand of Eq.\eqref{EQ:HallNormal} and Eq.\eqref{EQ:HallIntra} in (a) of Fig.\ref{FIG:integrandPlot} shows the change being limited to the elimination of the singularity at the Fermi surface similar to what we see for the occupation number.

\begin{widetext}
The change in the Hall conductivity due to infinitesimal pairing gaps has contributions from both the change in the occupation number $n_\pm ({\bf k})$ and the factor that was previously (in the Rashba metal) the Berry curvature 
\begin{equation}
\sigma^{SC}_{xy} - \sigma^{N}_{xy} 
\approx \frac{e^2}{2\pi}\int \frac{d^2 k}{(2\pi)^2} \sum_n [\{\Delta\mathcal{F}^{xy}_n\} n_n ({\bf k})+\mathcal{F}^{xy}_n\{\Delta n_n ({\bf k})\}]
\label{EQ:intraPairHall1}
\end{equation}
both terms in the order of $|\Delta|^2 \log |\Delta|$ for a small pairing gap at the Fermi surface; the first derivative with respect to the pairing gap vanishes. In the limit of small Rashba effect $\alpha \ll \sqrt{\mu / m^*}$ and Zeeman field $|h| \ll \mu$, the change in Hall conductivity comes mostly from $\Delta\mathcal{F}^{xy}_n$, giving us 
\begin{equation}
\sigma^{SC}_{xy} - \sigma^{N}_{xy} \approx -\frac{e^2}{32\pi}\frac{h}{\mu^2 m^* \alpha^2}\sum_s |\Delta_s|^2 \log \frac{4\sqrt{2} \alpha \sqrt{m^* \mu}}{|\Delta_s|}.
\label{EQ:intraPairHall2}
\end{equation}
\end{widetext}
While the factor of $\log |\Delta_-|$ would not be present in the $|\Delta_-|^2$ term for the topologically non-trivial superconducting phase, 
we do not expect this to have any substantial effect as $|\Delta_-| \ll |\Delta_+|$ is physically expected in the topologically non-trivial phase. Hence, we conclude that for the case of the purely intraband pairing there is no qualitative difference between the effect of superconductivity on the Hall conductivity between the topologically trivial and non-trivial intraband superconductivity.  This is consistent with the observation that the existence of the chiral Majorana edge state, the key feature of the topologically non-trivial superconductivity absent in the topologically trivial superconductivity, is irrelevant to the response to the electromagnetic field as the quasiparticle excitation of the Majorana edge state is charge neutral.

\subsection{Effect of interband pairing}

In this section, we will show how the effect of superconductivity on the Hall conductivity becomes qualitatively larger with interband pairing. Following the recent literature on the analysis of the topological superconductivity in the Rashba system \cite{Sau2010,Alicea2010,Oreg2010}, we return to the simplest form of the BCS pairing,
\begin{equation}
H_{pair} = |\Delta| \sum_{\bf k} c^\dagger_{{\bf k}\uparrow} c^\dagger_{-{\bf k}\downarrow} + {\rm h.c.},
\end{equation}
mentioned at the beginning of this section which gives us the full first-quantized Hamiltonian of
\begin{align}
\hat{U}^\dagger \hat{\mathcal{H}} \hat{U} =& \tau_3 (\xi_k - \sigma_3 d_k) - |\Delta| (-\tau_1 \hat{k}_y + \sigma_3 \tau_2 \hat{k}_x)\sqrt{1-d_z^2}\nonumber\\ 
&+ |\Delta| \sigma_2 \tau_2 \hat{d}_z \label{fullHam}
\end{align}
in the band basis as mentioned in the previous subsection. Note that, 
in addition to the interband pairings with the opposite chiralities, there is an interband pairing that is non-chiral \cite{Alicea2010}. This interband pairing at ${\bf k} = 0$ is responsible for shifting the quantum phase transition point from $\mu = |h|$ to $\mu = \sqrt{h^2 - |\Delta|^2}$. 
It is required in order to have a purely $s$-wave pairing because with the perpendicular Zeeman field, the ${\bf k}$ and $-{\bf k}$ states no longer has the opposite spins due to partial spin polarization. 

The interband pairing gives rise to contribution to the Hall conductivity that is not present in the normal state, {\it i.e.} not representable by (a) of Fig.\ref{FIG:HallDiagram}, due to the interband component of Green function being nonzero. This can be illustrated simply by the case with the infinitesimal interband pairing, for which we can set
\begin{align}
\hat{U}^\dagger \hat{\mathcal{H}}_0 \hat{U} =& \tau_3 (\xi_k - \sigma_3 d_k) - |\Delta| (-\tau_1 \hat{k}_y + \sigma_3 \tau_2 \hat{k}_x),\nonumber\\ 
\hat{U}^\dagger \delta\hat{\mathcal{H}} \hat{U} =& |\delta \Delta| \sigma_2 \tau_2 \hat{d}_z,
\end{align}
where $|\delta\Delta| \ll |\Delta|$. Note that with this model, the Green function has an interband component to the first order in the interband pairing,
\begin{equation}
\hat{U}^\dagger\delta \hat{G} ({\bf k},i\omega_n)  \hat{U} = (i\omega_n - \hat{U}^\dagger \hat{\mathcal{H}}_0 \hat{U}) \hat{U}^\dagger \delta\hat{\mathcal{H}} \hat{U} (i\omega_n - \hat{U}^\dagger \hat{\mathcal{H}}_0 \hat{U}),
\end{equation}
which has both the normal and anomalous part
\begin{align}
\delta\hat{g} =& \hat{f}\delta\hat{\mathcal{H}}\hat{g} + \hat{g}\delta\hat{\mathcal{H}}\hat{f},\nonumber\\
\delta\hat{f} =& \hat{f}\delta\hat{\mathcal{H}}\hat{f} + \hat{g}\delta\hat{\mathcal{H}}\hat{g},
\label{EQ:infInter1}
\end{align}
where $\hat{g}, \hat{f}$ is the normal and anomalous Green function for $\hat{\mathcal{H}}_0$. The result is analogous to the multiband chiral $p$-wave model of Sr$_2$RuO$_4$ \cite{Edward2012,Gradhand2013,Taylor2013} in having interband pairing turn on a process that contributes to Hall conductivity. However, while the results for Sr$_2$RuO$_4$ obtain Hall conductivity due to the anomalous interband Green function, we find that only the normal interband Green function contributes to the Hall conductivity,
\begin{align}
\delta \sigma_{xy} =& \frac{ie^2}{4\omega}\int \frac{d^2 k d\nu}{(2\pi)^3} \{{\rm tr}[\hat{v}_x \hat{g}({\bf k},i\nu+\omega)\hat{v}_y \delta\hat{g}({\bf k},i\nu)]\nonumber\\
&+ {\rm tr}[\hat{v}_x \delta\hat{g}({\bf k},i\nu+\omega)\hat{v}_y \hat{g}({\bf k},i\nu)]\}-(x \leftrightarrow y),
\label{EQ:infInter2}
\end{align}
through the process represented by (b) of Fig.\ref{FIG:HallDiagram}. Given that this process involves anomalous - that is, electron to hole or vice versa - propagation at some point, we expect it to maximize at the Fermi surfaces. Since with the nonzero interband pairing the Hall conductivity receives contribution from process different from that of the normal state, we can expect that this leads to qualitatively larger change in the Hall conductivity from its normal state value. Hence we expect the dependence of the Hall conductivity on the superconducting gap for the purely $s$-wave pairing to be qualitatively different from that for the purely intraband pairing. 

\begin{figure}[h]
\includegraphics[width=1.75in]{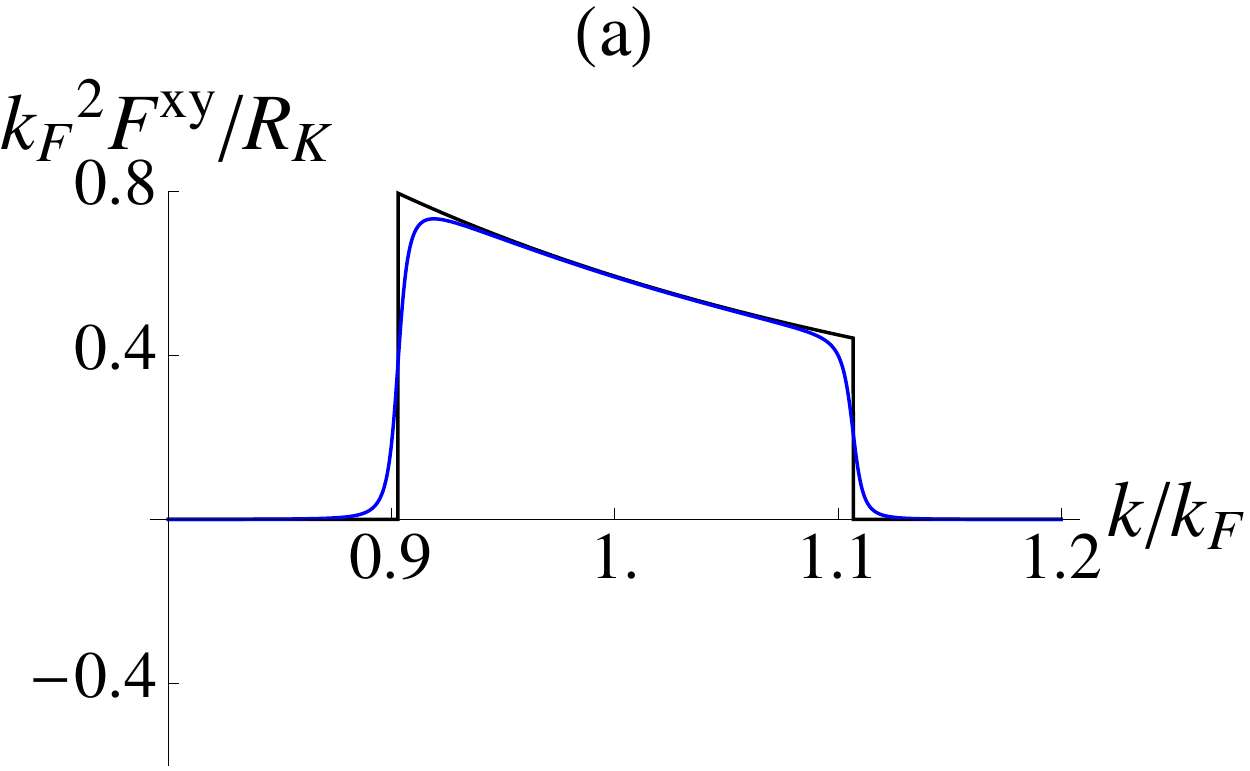}\includegraphics[width=1.75in]{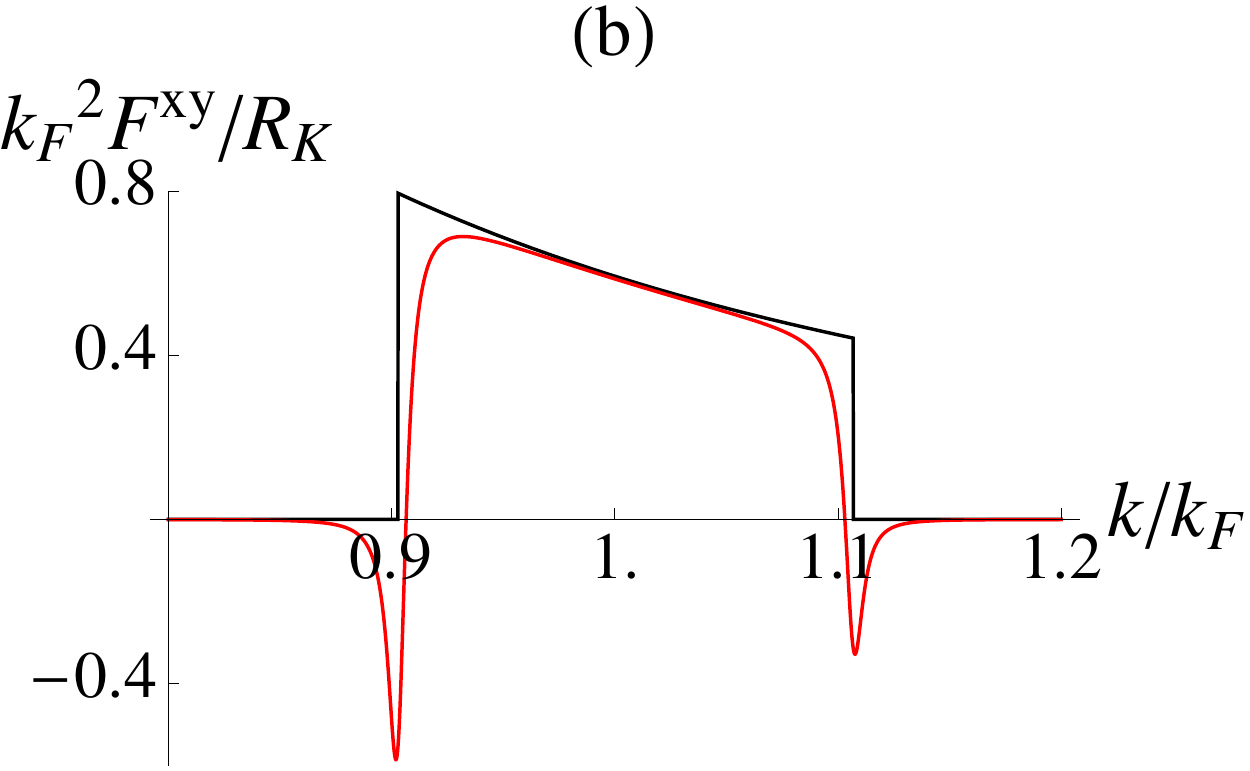}
\caption{Comparison between the integrand of the Hall conductivity of the normal state in Eq.\eqref{EQ:HallNormal} with that of (a) the purely intraband pairing superconductivity in Eq.\eqref{EQ:HallIntra} and (b) the purely $s$-wave superconductivity in Eq.\eqref{EQ:interHall}. 
We have defined $k_F \equiv \sqrt{2\mu m^*}$, $R_K = 2\pi/e^2$ (`the effective Berry curvature' $\mathcal{F}^{xy}$ normalized to give the normal state value for $|\Delta|=0$) and set for both (a) and (b) $m^* \alpha/k_F = 0.1$, $h/\alpha k_F = 0.2$  and $|\Delta|/\alpha k_F = 0.05$.}
\label{FIG:integrandPlot}
\end{figure}

\begin{widetext}
For the Hamiltonian of Eq.~\ref{fullHam}, we find that the Hall conductivity is 
\begin{align}
\sigma^{SC}_{xy} = \frac{e^2 \alpha^2 h}{2} \int\frac{d^2 k}{(2\pi)^2}\frac{1}{(\tilde{E}_{k+}+\tilde{E}_{k-})^3}&\left[\frac{4(\xi^2_k - d_k^2)^2}{\tilde{E}_{k+}^2 \tilde{E}_{k-}^2} \left(1-\frac{\xi_k^2 - d_k^2}{\tilde{E}_{k+} \tilde{E}_{k-}}\right)\right.\nonumber\\
&- 2|\Delta|^2 \left\{\frac{2(\xi^2_k - d_k^2)(3\xi^2_k- 3d_k^2 + 4\alpha^2 k^2 )}{\tilde{E}_{k+}^3 \tilde{E}_{k-}^3}+\frac{6(\xi_k^2 - d_k^2 + 2 \alpha^2 k^2)}{\tilde{E}_{k+}^2 \tilde{E}_{k-}^2}\right\}\nonumber\\
&\left.- 2|\Delta|^4 \left(\frac{6\xi^2_k - 6 d_k ^2 + 4\alpha^2 k^2}{\tilde{E}_{k+}^3 \tilde{E}_{k-}^3}+\frac{6}{\tilde{E}_{k+}^2 \tilde{E}_{k-}^2}\right)-4|\Delta|^6\frac{1}{\tilde{E}_{k+}^3 \tilde{E}_{k-}^3}\right]
\label{EQ:interHall}
\end{align}
where
\begin{equation}
%
\tilde{E}_{k\pm}^2 
=\xi_k^2 + d_k^2 + |\Delta|^2 \pm 2 \sqrt{\xi_k^2 d_k^2 +|\Delta|^2 h^2}
\end{equation}
(see Appendix \ref{formula} for derivation); note that in the $|\Delta| \to 0$ limit, Eq.\eqref{EQ:interHall} converges to the normal state Hall conductivity of Eq.\eqref{EQ:HallNormal} from the first term of the intergrand. The key change from the case of the purely intraband pairing is that the integrand of Eq.\eqref{EQ:interHall} actually reverses its sign at the Fermi surfaces $\xi_k \pm d_k = 0$. From the case of infinitesimal interband pairing examined above, we have seen that the interband propagation can lead to a large change at the Fermi surfaces . Comparison between the integrand of Eq.\eqref{EQ:interHall} and that of Eq.\eqref{EQ:HallNormal} shown in (b) of Fig.\ref{FIG:integrandPlot} clearly shows this striking change at the Fermi surface. 
The two plots of Fig.\ref{FIG:integrandPlot} strongly indicate that the change in the Hall conductivity is going to be qualitatively larger with the interband pairing than with the purely intraband pairing.

Indeed, from differentiating Eq.\eqref{EQ:interHall} with respect to $|\Delta|$, we find that the Hall conductivity is linear in the pairing gap in the $|\Delta| \to 0$ limit, which is a qualitatively stronger dependence than the $|\Delta|^2 \log |\Delta|$ dependence we find for the purely intraband pairing case of the previous section. We obtain 
\begin{equation}
\sigma^{SC}_{xy} - \sigma^{N}_{xy} \approx -\frac{3}{16} e^2 \alpha^2 h \left[\frac{k_{F+}}{v_{F+}}\frac{\alpha k_{F+}}{(h^2+\alpha^2 k_{F+}^2)^2}+\frac{k_{F-}}{v_{F-}}\frac{\alpha k_{F-}}{(h^2+\alpha^2 k_{F-}^2)^2}\right]|\Delta|
\label{EQ:interPairHall}
\end{equation}
($v_{F\pm}$'s are the velocity on the outer/inner Fermi surfaces) for the case where we have two Fermi surfaces. 
Given that the Landau-Ginzburg theory gives us  $\Delta \propto (T_c - T)^{1/2} $, the above result implies $\delta\sigma^{SC}_{xy} \sim (T_c - T)^{1/2}$ below $T_{c}$. 
\end{widetext}

Our results imply that for the purely $s$-wave pairing, the effect of pairing on the Hall conductivity does not change across the quantum phase transition at $h^2 = \mu^2 + |\Delta|^2$. 
To see this, note that in Eq.\eqref{EQ:interPairHall}, $(\partial/\partial |\Delta|)\sigma^{SC}_{xy}$ does not vanish on either side  of the transition. The only difference for the topologically non-trivial superconducting phase is that the $k_{F-}$ contribution of Eq.\eqref{EQ:interPairHall} vanishes, 
which leaves unchanged the contribution from the larger Fermi surface (the $k_{F+}$-dependent term) in Eq.\eqref{EQ:interPairHall}. This is fully in accord with our numerical results in (b) of Fig.\ref{FIG:integrandPlot} which show the $|\Delta|$-linear dependence to originate at the Fermi surfaces.




\section{Physics of interband pairing}

Our results indicate that the Hall conductivity provides a clear diagnostic for presence of interband pairing. It is therefore imperative to consider the circumstances under which the interband pairing could arise. 

In an intrinsic superconductivity with infinitesimally weak interaction, the Cooper pairing occurs between electrons of the same energy near the Fermi level, 
as the pairing of states at different energies 
cannot save as much energy.  Thus intrinsic pairing between different bands is unlikely to be energetically favorable unless the pairing is of the FFLO type 
with a nonzero center of mass momentum or the pairing interaction is strong compared to the gap between bands at the Fermi energy. 

Since we have not considered FFLO pairing,  the results of the previous section are relevant only when  the pairing interaction is strong  or when the 
superconductivity is induced through proximity effect and produces a pairing gap comparable to the splitting between the two bands $2d_k$. From both the analysis of the infinitesimal interband pairing in Eqs.\eqref{EQ:infInter1}, \eqref{EQ:infInter2} and the numerical results for the purely $s$-wave pairing in Fig.\ref{FIG:integrandPlot}, we see that the effect of interband pairing on the Hall conductivity is  significant mainly at the Fermi surfaces.

We emphasize that the physically relevant question is whether there can be interband pairing comparable to intraband pairing. It needs to be pointed out here that physically even the proximity effect will not induce a purely $s$-wave pairing. Tunneling between the Rashba system and the $s$-wave superconductor allows for spin-flip processes due to the spin-orbit coupling and may induce inverse-proximity effect resulting in a nonzero spin-triplet pairing correlation on the $s$-wave superconductor or a suppression of superconductivity in the $s$-wave superconductor. 
 
We conclude here that interband pairing is less likely in the topologically non-trivial superconducting phase 
which requires a single Fermi surface in the presence of the Zeeman field. 
We will show that usually the gap between the bands at the Fermi surface are larger than in the case of a single Fermi surface than in the case of Fermi surfaces in both bands. This in turn makes inter-band paring more energetically unfavorable in the case of topologically non-trivial superconductivity than in the case of the trivial superconducting phase. To illustrate this point, it is useful to examine certain specific ranges of the parameters.

Consider the special point $\mu=0$ at which we have a topologically non-trivial superconductivity for any $h^2 > |\Delta|^2 > 0$. At the limit of $h \to 0$ the Fermi surface would be at $k_{F+} = 2m^*\alpha$, and hence the band splitting at the Fermi surface would be $2\alpha k_{F+} = 4m^*\alpha^2$. 
In the $h \ll m^* \alpha^2$ limit, the band splitting at the Fermi surface would be much larger then $h$ and hence also the pairing gap $|\Delta|$. 
Possibility for the interband pairing only exists in the experimentally challenging regime of $|h| \gtrsim |\Delta| \gg m^*\alpha^2$.

There are less constraints in the  topologically trivial superconducting phase, since for a fixed $h, |\Delta|$,  the chemical potential $\mu$ needs not be fine-tuned. One possible scenario is in the limit of small spin-orbit coupling and Zeeman energy, {\it i.e.} $\mu \gg |h|, m\alpha^2$. In this case there is no restriction against the band splitting $2d_k \approx 2\sqrt{h^2 + 2\mu m^* \alpha^2}$ becoming comparable to $|\Delta|$, and 
the interband pairing on the both Fermi surfaces will give us
\begin{equation}
\sigma^{SC}_{xy} - \sigma^{N}_{xy} \approx -\frac{3}{4\sqrt{2}} e^2 \frac{\mu^{1/2} (m^* \alpha^2)^{3/2} h |\Delta|}{(h^2 + 2\mu m^* \alpha^2)^2}. 
\end{equation}
Another scenario is for the case where we are in the trivial superconducting phase, yet close to the quantum phase transition, {\it e.g.}, $0<\mu-|h|\ll \mu$. For $|\Delta|$ comparable to $|h|$, there can be significant interband pairing at the smaller Fermi surface 
giving us
\begin{equation}
\sigma^{SC}_{xy} - \sigma^{N}_{xy} \approx -\frac{3}{16} e^2 \frac{\alpha k_{F-}}{h^2} |\Delta|. 
\end{equation}

\section{Conclusion and Discussion}

 We have studied the intrinsic Hall conductivity in the normal and superconducting phases of the Rashba system under perpendicular magnetic field. In this system, the normal state itself has broken time-reversal symmetry and a non-zero intrinsic Hall conductivity, in contrast to Sr$_2$RuO$_4$, where the normal system is time-reversal invariant and Cooper pairing breaks time-reversal symmetry. We have compared two cases for this system, one where the Cooper pairing is exclusively intra-band and the other where we allowed for the inter-band pairing as well; both cases have the same topologically trivial and non-trivial superconducting phases and the identical class of the quantum phase transition between them tuned by the chemical potential $\mu$. For either case, we find no qualitative difference between the Hall conductivity in the topologically trivial and non-trivial superconducting phases. On the other hand we find that between the case with zero and non-zero inter-band pairing, the dependence of the Hall conductivity on the pairing gap is qualitatively different, with the non-zero inter-band pairing case having a Hall conductivity linear in the pairing gap.

Experimentally, our result suggests that while the observation of the linear dependence of intrinsic Hall conductivity on the pairing gap is more likely to be associated with the topologically trivial superconductivity, if not quite ruling out the topologically non-trivial superconductivity. This is because the interband pairing in the topologically non-trivial superconducting phase requires either a very strong Zeeman field or a very low electron density and therefore would be very difficult to realize, whereas in the case of the topologically trivial superconducting phase one merely needs a sufficiently weak spin-orbit coupling.

Our Hall conductivity results on the Rashba system are consistent with those on the chiral $p$-wave superconductor. While the chiral $p$-wave superconductor always breaks time-reversal symmetry, its topology depends on the number of pockets crossing the Fermi level, being is non-trivial (trivial) for odd (even) number of pockets. A non-zero Hall conductivity in an impurity-free chiral $p$-wave superconductor requires inter-band pairing regardless of whether the superconductivity is topologically trivial or non-trivial. Our results are consistent in both aspects : on the importance of the inter-band pairing on the Hall conductivity and on the absence of any qualitative dependence on the topology of the superconducting phase.

The non-quantization of the Hall conductivity in superconductors is another consequence of the sharply different electromagnetic response of superconductors and insulators. 
The quantization in insulators can be explained by an argument which relies on an adiabatic  insertion of  $h/e$ flux through a ring in the Corbino geometry. 
Since the inserted flux can be ``gauged away'', the system must return to its initial state 
with a possible transport of an integer number of electrons 
from the inner edge to the outer edge \cite{Laughlin1981,Niu1985}. However, in superconductor, flux insertion leads to Meissner screening, and eventually, by the time a $h/2e$ flux is inserted, the phase slip by which the superconductor acquires the $2\pi$ phase winding around the hole occurs in the superconductor to reduce the kinetic energy. Since the $2\pi$ phase winding cannot occur adiabatically, the $h/e$ flux cannot be adiabatically inserted in a superconductor and the quantization argument does not apply. 

We would like to thank Catherine Kallin, Sudip Chakravarty and Srinivas Raghu for useful discussions and suggestions. This work has been supported by the UCLA Startup Funds (RR, SBC) and the Institute for Basic Science in Korea through the Young Scientist grant (SBC).

\appendix

\begin{widetext}
\section{Hall conductivity for the purely $s$-wave pairing Rashba system}
\label{formula}

The fact that the model is {\it not} block-diagonal in the band basis introduces a little complication to the Green's function:
\begin{align}
\hat{G}({\bf k},i\omega_n) = [i\omega_n - \hat{\mathcal{H}}({\bf k})]^{-1} = \frac{\hat{g}_0 (k,i\omega_n) -\alpha\sum_{i=1,2}\sigma_i [k_x \hat{g}_{ix}(k,i\omega_n) +k_y \hat{g}_{iy}(k,i\omega_n)]-\sigma_3 h\hat{g}_3(k,i\omega_n)}{(\omega_n^2+\tilde{E}_{k+}^2)(\omega_n^2+\tilde{E}_{k-}^2)}.
\end{align}
where 
\begin{align}
\tilde{E}_{k\pm}^2 =&  \xi_k^2 + d_k^2 + |\Delta|^2 \pm 2 \sqrt{\xi_k^2 d_k^2 +|\Delta|^2 h^2},\nonumber\\
\hat{g}_0(k,i\omega_n) = & -i\omega_n(\omega_n^2 + \xi_k^2 + d_k^2 + |\Delta|^2) -\tau_3 \xi_k(\omega_n^2 + \xi_k^2 - d_k^2 + |\Delta|^2)\nonumber\\ 
&-\tau_1|\Delta|(\omega_n^2 + \xi_k^2 + \alpha ^2 k^2 - h^2  + |\Delta|^2),\nonumber\\
\hat{g}_{2x}(k,i\omega_n) =  -\hat{g}_{1y}(k,i\omega_n) =&  2i \omega_n\xi_k - \tau_3 (\omega_n^2-\xi^2_k + d_k^2 +|\Delta|^2) + \tau_1 2\xi_k |\Delta|,\nonumber\\
\hat{g}_{1x}(k,i\omega_n) =  \hat{g}_{2y}(k,i\omega_n) =& -\tau_2 2 h |\Delta|,\nonumber\\
\hat{g}_3(k,i\omega_n) = & (\omega_n^2 - \xi_k^2 + d_k^2  - |\Delta|^2) -\tau_3 2i\omega_n\xi_k - \tau_1 2i\omega_n |\Delta|
\end{align}
(note that all $\hat{g}$'s are independent of spin and ${\bf \hat{k}}$). However, the above formula does permit writing down in a relatively simple form the optical Hall conductivity at $T=0$ after tracing over the spins:
\begin{align}
\sigma_{xy}(\omega) =& e^2 \alpha^2 h \int\frac{d^2 k d\nu}{(2\pi)^3} \frac{{\rm tr}[\hat{g}_0(k,i\nu)\hat{g}_3(k,i\nu+\omega)-\hat{g}_3(k,i\nu)\hat{g}_0(k,i\nu+\omega)]}{\omega(\nu^2+\tilde{E}_{k+}^2)(\nu^2+\tilde{E}_{k-}^2)[(\nu-i\omega)^2+\tilde{E}_{k+}^2][(\nu-i\omega)^2+\tilde{E}_{k-}^2]}
\end{align}
(tr here is only over the electron and hole), from which we obtain Eq.\eqref{EQ:interHall} in the $\omega \to 0$ limit.  

Taking the first derivative of Eq.\eqref{EQ:interHall} with respect to the pairing gap $|\Delta|$ in the $|\Delta| \to 0$ limit gives us 
\begin{align}
\left.\frac{\partial}{\partial |\Delta|}\right\vert_{|\Delta| \to 0} \sigma^{SC}_{xy} =& -12e^2 \alpha^2 h \int\frac{d^2 k}{(2\pi)^2}\lim_{|\Delta| \to 0} \frac{\xi_k^2 -d_k^2 + 2\alpha^2 k^2}{(\tilde{E}_{k+}+\tilde{E}_{k-})^3 \tilde{E}_{k+}^2}\frac{\partial}{\partial |\Delta|}\frac{|\Delta|^2}{\tilde{E}_{k-}^2}\nonumber\\
=& -12 e^2 \alpha^2 h \lim_{|\Delta| \to 0} |\Delta| \int\frac{d^2 k}{(2\pi)^2} \frac{\xi_k^2 -d_k^2 + 2\alpha^2 k^2}{(|\xi_k+d_k|+|\xi_k-d_k|)^3 (|\xi_k|+d_k)^2}\nonumber\\
&\times\left[\frac{1}{(|\xi_k|-d_k)^2 + |\Delta|^2(1-h^2/|\xi_k d_k|)}-\frac{|\Delta|^2(1-h^2/|\xi_k d_k|)}{\{(|\xi_k|-d_k)^2 + |\Delta|^2(1-h^2/|\xi_k d_k|)\}^2}\right]\nonumber\\
\approx& -\frac{3 e^2 \alpha^2 h}{8\pi}\lim_{|\Delta| \to 0}|\Delta|\int dk \sum_{s=\pm}\frac{k_{Fs}(\alpha^2 k_{Fs}^2)}{(h^2+\alpha^2 k_{Fs}^2)^{5/2}}\left[\frac{1}{v_{Fs}^2(k-k_{Fs})^2 + |\Delta_s|^2}-\frac{|\Delta_s|^2}{\{v_{Fs}^2(k-k_{Fs})^2 + |\Delta_s|^2\}^2}\right]\nonumber\\
=& -\frac{3}{16} e^2 \alpha^2 h \left[\frac{k_{F+}}{v_{F+}}\frac{\alpha k_{F+}}{(h^2+\alpha^2 k_{F+}^2)^2}+\frac{k_{F-}}{v_{F-}}\frac{\alpha k_{F-}}{(h^2+\alpha^2 k_{F-}^2)^2}\right],
\label{EQ:interPairHall2}
\end{align}
where $|\tilde{\Delta}_\pm|^2 \equiv  E_{k\pm}^2 - (\xi_k \mp d_k)^2 \approx |\Delta|^2\alpha^2 k_{F\pm}^2 /(h^2 + \alpha^2 k_{F\pm}^2)$. 

\end{widetext}

\bibliography{Hall}

\begin{thebibliography}{30}%
\makeatletter
\providecommand \@ifxundefined [1]{%
 \@ifx{#1\undefined}
}%
\providecommand \@ifnum [1]{%
 \ifnum #1\expandafter \@firstoftwo
 \else \expandafter \@secondoftwo
 \fi
}%
\providecommand \@ifx [1]{%
 \ifx #1\expandafter \@firstoftwo
 \else \expandafter \@secondoftwo
 \fi
}%
\providecommand \natexlab [1]{#1}%
\providecommand \enquote  [1]{``#1''}%
\providecommand \bibnamefont  [1]{#1}%
\providecommand \bibfnamefont [1]{#1}%
\providecommand \citenamefont [1]{#1}%
\providecommand \href@noop [0]{\@secondoftwo}%
\providecommand \href [0]{\begingroup \@sanitize@url \@href}%
\providecommand \@href[1]{\@@startlink{#1}\@@href}%
\providecommand \@@href[1]{\endgroup#1\@@endlink}%
\providecommand \@sanitize@url [0]{\catcode `\\12\catcode `\$12\catcode
  `\&12\catcode `\#12\catcode `\^12\catcode `\_12\catcode `\%12\relax}%
\providecommand \@@startlink[1]{}%
\providecommand \@@endlink[0]{}%
\providecommand \url  [0]{\begingroup\@sanitize@url \@url }%
\providecommand \@url [1]{\endgroup\@href {#1}{\urlprefix }}%
\providecommand \urlprefix  [0]{URL }%
\providecommand \Eprint [0]{\href }%
\providecommand \doibase [0]{http://dx.doi.org/}%
\providecommand \selectlanguage [0]{\@gobble}%
\providecommand \bibinfo  [0]{\@secondoftwo}%
\providecommand \bibfield  [0]{\@secondoftwo}%
\providecommand \translation [1]{[#1]}%
\providecommand \BibitemOpen [0]{}%
\providecommand \bibitemStop [0]{}%
\providecommand \bibitemNoStop [0]{.\EOS\space}%
\providecommand \EOS [0]{\spacefactor3000\relax}%
\providecommand \BibitemShut  [1]{\csname bibitem#1\endcsname}%
\let\auto@bib@innerbib\@empty
\bibitem [{\citenamefont {Laughlin}(1981)}]{Laughlin1981}%
  \BibitemOpen
  \bibfield  {author} {\bibinfo {author} {\bibfnamefont {R.~B.}\ \bibnamefont
  {Laughlin}},\ }\href {\doibase 10.1103/PhysRevB.23.5632} {\bibfield
  {journal} {\bibinfo  {journal} {Phys. Rev. B}\ }\textbf {\bibinfo {volume}
  {23}},\ \bibinfo {pages} {5632} (\bibinfo {year} {1981})}\BibitemShut
  {NoStop}%
\bibitem [{\citenamefont {Thouless}\ \emph {et~al.}(1982)\citenamefont
  {Thouless}, \citenamefont {Kohmoto}, \citenamefont {Nightingale},\ and\
  \citenamefont {den Nijs}}]{Thouless1982}%
  \BibitemOpen
  \bibfield  {author} {\bibinfo {author} {\bibfnamefont {D.~J.}\ \bibnamefont
  {Thouless}}, \bibinfo {author} {\bibfnamefont {M.}~\bibnamefont {Kohmoto}},
  \bibinfo {author} {\bibfnamefont {M.~P.}\ \bibnamefont {Nightingale}}, \ and\
  \bibinfo {author} {\bibfnamefont {M.}~\bibnamefont {den Nijs}},\ }\href
  {\doibase 10.1103/PhysRevLett.49.405} {\bibfield  {journal} {\bibinfo
  {journal} {Phys. Rev. Lett.}\ }\textbf {\bibinfo {volume} {49}},\ \bibinfo
  {pages} {405} (\bibinfo {year} {1982})}\BibitemShut {NoStop}%
\bibitem [{\citenamefont {Haldane}(1988)}]{Haldane1988}%
  \BibitemOpen
  \bibfield  {author} {\bibinfo {author} {\bibfnamefont {F.~D.~M.}\
  \bibnamefont {Haldane}},\ }\href {\doibase 10.1103/PhysRevLett.61.2015}
  {\bibfield  {journal} {\bibinfo  {journal} {Phys. Rev. Lett.}\ }\textbf
  {\bibinfo {volume} {61}},\ \bibinfo {pages} {2015} (\bibinfo {year}
  {1988})}\BibitemShut {NoStop}%
\bibitem [{\citenamefont {Liu}\ \emph {et~al.}(2008)\citenamefont {Liu},
  \citenamefont {Qi}, \citenamefont {Dai}, \citenamefont {Fang},\ and\
  \citenamefont {Zhang}}]{Liu2008}%
  \BibitemOpen
  \bibfield  {author} {\bibinfo {author} {\bibfnamefont {C.-X.}\ \bibnamefont
  {Liu}}, \bibinfo {author} {\bibfnamefont {X.-L.}\ \bibnamefont {Qi}},
  \bibinfo {author} {\bibfnamefont {X.}~\bibnamefont {Dai}}, \bibinfo {author}
  {\bibfnamefont {Z.}~\bibnamefont {Fang}}, \ and\ \bibinfo {author}
  {\bibfnamefont {S.-C.}\ \bibnamefont {Zhang}},\ }\href {\doibase
  10.1103/PhysRevLett.101.146802} {\bibfield  {journal} {\bibinfo  {journal}
  {Phys. Rev. Lett.}\ }\textbf {\bibinfo {volume} {101}},\ \bibinfo {pages}
  {146802} (\bibinfo {year} {2008})}\BibitemShut {NoStop}%
\bibitem [{\citenamefont {Chang}\ \emph {et~al.}(2013)\citenamefont {Chang},
  \citenamefont {Zhang}, \citenamefont {Feng}, \citenamefont {Shen},
  \citenamefont {Zhang}, \citenamefont {Guo}, \citenamefont {Li}, \citenamefont
  {Ou}, \citenamefont {Wei}, \citenamefont {Wang}, \citenamefont {Ji},
  \citenamefont {Feng}, \citenamefont {Ji}, \citenamefont {Chen}, \citenamefont
  {Jia}, \citenamefont {Dai}, \citenamefont {Fang}, \citenamefont {Zhang},
  \citenamefont {He}, \citenamefont {Wang}, \citenamefont {Lu}, \citenamefont
  {Ma},\ and\ \citenamefont {Xue}}]{Chang2013}%
  \BibitemOpen
  \bibfield  {author} {\bibinfo {author} {\bibfnamefont {C.-Z.}\ \bibnamefont
  {Chang}}, \bibinfo {author} {\bibfnamefont {J.}~\bibnamefont {Zhang}},
  \bibinfo {author} {\bibfnamefont {X.}~\bibnamefont {Feng}}, \bibinfo {author}
  {\bibfnamefont {J.}~\bibnamefont {Shen}}, \bibinfo {author} {\bibfnamefont
  {Z.}~\bibnamefont {Zhang}}, \bibinfo {author} {\bibfnamefont
  {M.}~\bibnamefont {Guo}}, \bibinfo {author} {\bibfnamefont {K.}~\bibnamefont
  {Li}}, \bibinfo {author} {\bibfnamefont {Y.}~\bibnamefont {Ou}}, \bibinfo
  {author} {\bibfnamefont {P.}~\bibnamefont {Wei}}, \bibinfo {author}
  {\bibfnamefont {L.-L.}\ \bibnamefont {Wang}}, \bibinfo {author}
  {\bibfnamefont {Z.-Q.}\ \bibnamefont {Ji}}, \bibinfo {author} {\bibfnamefont
  {Y.}~\bibnamefont {Feng}}, \bibinfo {author} {\bibfnamefont {S.}~\bibnamefont
  {Ji}}, \bibinfo {author} {\bibfnamefont {X.}~\bibnamefont {Chen}}, \bibinfo
  {author} {\bibfnamefont {J.}~\bibnamefont {Jia}}, \bibinfo {author}
  {\bibfnamefont {X.}~\bibnamefont {Dai}}, \bibinfo {author} {\bibfnamefont
  {Z.}~\bibnamefont {Fang}}, \bibinfo {author} {\bibfnamefont {S.-C.}\
  \bibnamefont {Zhang}}, \bibinfo {author} {\bibfnamefont {K.}~\bibnamefont
  {He}}, \bibinfo {author} {\bibfnamefont {Y.}~\bibnamefont {Wang}}, \bibinfo
  {author} {\bibfnamefont {L.}~\bibnamefont {Lu}}, \bibinfo {author}
  {\bibfnamefont {X.-C.}\ \bibnamefont {Ma}}, \ and\ \bibinfo {author}
  {\bibfnamefont {Q.-K.}\ \bibnamefont {Xue}},\ }\href {\doibase
  10.1126/science.1234414} {\bibfield  {journal} {\bibinfo  {journal}
  {Science}\ }\textbf {\bibinfo {volume} {340}},\ \bibinfo {pages} {167}
  (\bibinfo {year} {2013})}\BibitemShut {NoStop}%
\bibitem [{\citenamefont {Karplus}\ and\ \citenamefont
  {Luttinger}(1954)}]{Karplus1954}%
  \BibitemOpen
  \bibfield  {author} {\bibinfo {author} {\bibfnamefont {R.}~\bibnamefont
  {Karplus}}\ and\ \bibinfo {author} {\bibfnamefont {J.~M.}\ \bibnamefont
  {Luttinger}},\ }\href {\doibase 10.1103/PhysRev.95.1154} {\bibfield
  {journal} {\bibinfo  {journal} {Phys. Rev.}\ }\textbf {\bibinfo {volume}
  {95}},\ \bibinfo {pages} {1154} (\bibinfo {year} {1954})}\BibitemShut
  {NoStop}%
\bibitem [{\citenamefont {Haldane}(2004)}]{Haldane2004}%
  \BibitemOpen
  \bibfield  {author} {\bibinfo {author} {\bibfnamefont {F.~D.~M.}\
  \bibnamefont {Haldane}},\ }\href {\doibase 10.1103/PhysRevLett.93.206602}
  {\bibfield  {journal} {\bibinfo  {journal} {Phys. Rev. Lett.}\ }\textbf
  {\bibinfo {volume} {93}},\ \bibinfo {pages} {206602} (\bibinfo {year}
  {2004})}\BibitemShut {NoStop}%
\bibitem [{\citenamefont {Xia}\ \emph {et~al.}(2006)\citenamefont {Xia},
  \citenamefont {Maeno}, \citenamefont {Beyersdorf}, \citenamefont {Fejer},\
  and\ \citenamefont {Kapitulnik}}]{Xia2006}%
  \BibitemOpen
  \bibfield  {author} {\bibinfo {author} {\bibfnamefont {J.}~\bibnamefont
  {Xia}}, \bibinfo {author} {\bibfnamefont {Y.}~\bibnamefont {Maeno}}, \bibinfo
  {author} {\bibfnamefont {P.~T.}\ \bibnamefont {Beyersdorf}}, \bibinfo
  {author} {\bibfnamefont {M.~M.}\ \bibnamefont {Fejer}}, \ and\ \bibinfo
  {author} {\bibfnamefont {A.}~\bibnamefont {Kapitulnik}},\ }\href {\doibase
  10.1103/PhysRevLett.97.167002} {\bibfield  {journal} {\bibinfo  {journal}
  {Phys. Rev. Lett.}\ }\textbf {\bibinfo {volume} {97}},\ \bibinfo {pages}
  {167002} (\bibinfo {year} {2006})}\BibitemShut {NoStop}%
\bibitem [{\citenamefont {{Schemm}}(2014)}]{Schemm2014}%
  \BibitemOpen
  \bibfield  {author} {\bibinfo {author} {\bibfnamefont {E.}~\bibnamefont
  {{Schemm}}},\ }in\ \href@noop {} {\emph {\bibinfo {booktitle} {APS Meeting
  Abstracts}}}\ (\bibinfo {year} {2014})\ p.\ \bibinfo {pages}
  {M48002}\BibitemShut {NoStop}%
\bibitem [{\citenamefont {{Schemm}}\ \emph {et~al.}(2013)\citenamefont
  {{Schemm}}, \citenamefont {{Karapetyan}}, \citenamefont {{Bauer}},\ and\
  \citenamefont {{Kapitulnik}}}]{Schemm2013}%
  \BibitemOpen
  \bibfield  {author} {\bibinfo {author} {\bibfnamefont {E.}~\bibnamefont
  {{Schemm}}}, \bibinfo {author} {\bibfnamefont {H.}~\bibnamefont
  {{Karapetyan}}}, \bibinfo {author} {\bibfnamefont {E.}~\bibnamefont
  {{Bauer}}}, \ and\ \bibinfo {author} {\bibfnamefont {A.}~\bibnamefont
  {{Kapitulnik}}},\ }in\ \href@noop {} {\emph {\bibinfo {booktitle} {APS
  Meeting Abstracts}}}\ (\bibinfo {year} {2013})\ p.\ \bibinfo {pages}
  {B35007}\BibitemShut {NoStop}%
\bibitem [{\citenamefont {Mackenzie}\ and\ \citenamefont
  {Maeno}(2003)}]{Mackenzie2003}%
  \BibitemOpen
  \bibfield  {author} {\bibinfo {author} {\bibfnamefont {A.~P.}\ \bibnamefont
  {Mackenzie}}\ and\ \bibinfo {author} {\bibfnamefont {Y.}~\bibnamefont
  {Maeno}},\ }\href {\doibase 10.1103/RevModPhys.75.657} {\bibfield  {journal}
  {\bibinfo  {journal} {Rev. Mod. Phys.}\ }\textbf {\bibinfo {volume} {75}},\
  \bibinfo {pages} {657} (\bibinfo {year} {2003})}\BibitemShut {NoStop}%
\bibitem [{\citenamefont {Luke}\ \emph {et~al.}(1998)\citenamefont {Luke},
  \citenamefont {Fudamoto}, \citenamefont {Kojima}, \citenamefont {Larkin},
  \citenamefont {Merrin}, \citenamefont {Nachumi}, \citenamefont {Uemura},
  \citenamefont {Maeno}, \citenamefont {Mao}, \citenamefont {Mori},
  \citenamefont {Nakamura},\ and\ \citenamefont {Sigrist}}]{Luke1998}%
  \BibitemOpen
  \bibfield  {author} {\bibinfo {author} {\bibfnamefont {G.~M.}\ \bibnamefont
  {Luke}}, \bibinfo {author} {\bibfnamefont {Y.}~\bibnamefont {Fudamoto}},
  \bibinfo {author} {\bibfnamefont {K.~M.}\ \bibnamefont {Kojima}}, \bibinfo
  {author} {\bibfnamefont {M.~I.}\ \bibnamefont {Larkin}}, \bibinfo {author}
  {\bibfnamefont {J.}~\bibnamefont {Merrin}}, \bibinfo {author} {\bibfnamefont
  {B.}~\bibnamefont {Nachumi}}, \bibinfo {author} {\bibfnamefont {Y.~J.}\
  \bibnamefont {Uemura}}, \bibinfo {author} {\bibfnamefont {Y.}~\bibnamefont
  {Maeno}}, \bibinfo {author} {\bibfnamefont {Z.~Q.}\ \bibnamefont {Mao}},
  \bibinfo {author} {\bibfnamefont {Y.}~\bibnamefont {Mori}}, \bibinfo {author}
  {\bibfnamefont {H.}~\bibnamefont {Nakamura}}, \ and\ \bibinfo {author}
  {\bibfnamefont {M.}~\bibnamefont {Sigrist}},\ }\href {\doibase 10.1038/29038}
  {\bibfield  {journal} {\bibinfo  {journal} {Nature}\ }\textbf {\bibinfo
  {volume} {394}},\ \bibinfo {pages} {558} (\bibinfo {year}
  {1998})}\BibitemShut {NoStop}%
\bibitem [{\citenamefont {Lutchyn}\ \emph {et~al.}(2008)\citenamefont
  {Lutchyn}, \citenamefont {Nagornykh},\ and\ \citenamefont
  {Yakovenko}}]{Lutchyn2008}%
  \BibitemOpen
  \bibfield  {author} {\bibinfo {author} {\bibfnamefont {R.~M.}\ \bibnamefont
  {Lutchyn}}, \bibinfo {author} {\bibfnamefont {P.}~\bibnamefont {Nagornykh}},
  \ and\ \bibinfo {author} {\bibfnamefont {V.~M.}\ \bibnamefont {Yakovenko}},\
  }\href {\doibase 10.1103/PhysRevB.77.144516} {\bibfield  {journal} {\bibinfo
  {journal} {Phys. Rev. B}\ }\textbf {\bibinfo {volume} {77}},\ \bibinfo
  {pages} {144516} (\bibinfo {year} {2008})}\BibitemShut {NoStop}%
\bibitem [{\citenamefont {Roy}\ and\ \citenamefont {Kallin}(2008)}]{Roy2008}%
  \BibitemOpen
  \bibfield  {author} {\bibinfo {author} {\bibfnamefont {R.}~\bibnamefont
  {Roy}}\ and\ \bibinfo {author} {\bibfnamefont {C.}~\bibnamefont {Kallin}},\
  }\href {\doibase 10.1103/PhysRevB.77.174513} {\bibfield  {journal} {\bibinfo
  {journal} {Phys. Rev. B}\ }\textbf {\bibinfo {volume} {77}},\ \bibinfo
  {pages} {174513} (\bibinfo {year} {2008})}\BibitemShut {NoStop}%
\bibitem [{\citenamefont {Taylor}\ and\ \citenamefont
  {Kallin}(2012)}]{Edward2012}%
  \BibitemOpen
  \bibfield  {author} {\bibinfo {author} {\bibfnamefont {E.}~\bibnamefont
  {Taylor}}\ and\ \bibinfo {author} {\bibfnamefont {C.}~\bibnamefont
  {Kallin}},\ }\href {\doibase 10.1103/PhysRevLett.108.157001} {\bibfield
  {journal} {\bibinfo  {journal} {Phys. Rev. Lett.}\ }\textbf {\bibinfo
  {volume} {108}},\ \bibinfo {pages} {157001} (\bibinfo {year}
  {2012})}\BibitemShut {NoStop}%
\bibitem [{\citenamefont {Gradhand}\ \emph {et~al.}(2013)\citenamefont
  {Gradhand}, \citenamefont {Wysokinski}, \citenamefont {Annett},\ and\
  \citenamefont {Gy\"orffy}}]{Gradhand2013}%
  \BibitemOpen
  \bibfield  {author} {\bibinfo {author} {\bibfnamefont {M.}~\bibnamefont
  {Gradhand}}, \bibinfo {author} {\bibfnamefont {K.~I.}\ \bibnamefont
  {Wysokinski}}, \bibinfo {author} {\bibfnamefont {J.~F.}\ \bibnamefont
  {Annett}}, \ and\ \bibinfo {author} {\bibfnamefont {B.~L.}\ \bibnamefont
  {Gy\"orffy}},\ }\href {\doibase 10.1103/PhysRevB.88.094504} {\bibfield
  {journal} {\bibinfo  {journal} {Phys. Rev. B}\ }\textbf {\bibinfo {volume}
  {88}},\ \bibinfo {pages} {094504} (\bibinfo {year} {2013})}\BibitemShut
  {NoStop}%
\bibitem [{\citenamefont {Taylor}\ and\ \citenamefont
  {Kallin}(2013)}]{Taylor2013}%
  \BibitemOpen
  \bibfield  {author} {\bibinfo {author} {\bibfnamefont {E.}~\bibnamefont
  {Taylor}}\ and\ \bibinfo {author} {\bibfnamefont {C.}~\bibnamefont
  {Kallin}},\ }\href@noop {} {\bibfield  {journal} {\bibinfo  {journal} {J.
  Phys.: Conf. Ser.}\ }\textbf {\bibinfo {volume} {449}},\ \bibinfo {pages}
  {012036} (\bibinfo {year} {2013})}\BibitemShut {NoStop}%
\bibitem [{\citenamefont {Sau}\ \emph {et~al.}(2010)\citenamefont {Sau},
  \citenamefont {Lutchyn}, \citenamefont {Tewari},\ and\ \citenamefont
  {Das~Sarma}}]{Sau2010}%
  \BibitemOpen
  \bibfield  {author} {\bibinfo {author} {\bibfnamefont {J.~D.}\ \bibnamefont
  {Sau}}, \bibinfo {author} {\bibfnamefont {R.~M.}\ \bibnamefont {Lutchyn}},
  \bibinfo {author} {\bibfnamefont {S.}~\bibnamefont {Tewari}}, \ and\ \bibinfo
  {author} {\bibfnamefont {S.}~\bibnamefont {Das~Sarma}},\ }\href {\doibase
  10.1103/PhysRevLett.104.040502} {\bibfield  {journal} {\bibinfo  {journal}
  {Phys. Rev. Lett.}\ }\textbf {\bibinfo {volume} {104}},\ \bibinfo {pages}
  {040502} (\bibinfo {year} {2010})}\BibitemShut {NoStop}%
\bibitem [{\citenamefont {Alicea}(2010)}]{Alicea2010}%
  \BibitemOpen
  \bibfield  {author} {\bibinfo {author} {\bibfnamefont {J.}~\bibnamefont
  {Alicea}},\ }\href {\doibase 10.1103/PhysRevB.81.125318} {\bibfield
  {journal} {\bibinfo  {journal} {Phys. Rev. B}\ }\textbf {\bibinfo {volume}
  {81}},\ \bibinfo {pages} {125318} (\bibinfo {year} {2010})}\BibitemShut
  {NoStop}%
\bibitem [{\citenamefont {Oreg}\ \emph {et~al.}(2010)\citenamefont {Oreg},
  \citenamefont {Refael},\ and\ \citenamefont {von Oppen}}]{Oreg2010}%
  \BibitemOpen
  \bibfield  {author} {\bibinfo {author} {\bibfnamefont {Y.}~\bibnamefont
  {Oreg}}, \bibinfo {author} {\bibfnamefont {G.}~\bibnamefont {Refael}}, \ and\
  \bibinfo {author} {\bibfnamefont {F.}~\bibnamefont {von Oppen}},\ }\href
  {\doibase 10.1103/PhysRevLett.105.177002} {\bibfield  {journal} {\bibinfo
  {journal} {Phys. Rev. Lett.}\ }\textbf {\bibinfo {volume} {105}},\ \bibinfo
  {pages} {177002} (\bibinfo {year} {2010})}\BibitemShut {NoStop}%
\bibitem [{\citenamefont {Mourik}\ \emph {et~al.}(2012)\citenamefont {Mourik},
  \citenamefont {Zuo}, \citenamefont {Frolov}, \citenamefont {Plissard},
  \citenamefont {Bakkers},\ and\ \citenamefont {Kouwenhoven}}]{Mourik2012}%
  \BibitemOpen
  \bibfield  {author} {\bibinfo {author} {\bibfnamefont {V.}~\bibnamefont
  {Mourik}}, \bibinfo {author} {\bibfnamefont {K.}~\bibnamefont {Zuo}},
  \bibinfo {author} {\bibfnamefont {S.~M.}\ \bibnamefont {Frolov}}, \bibinfo
  {author} {\bibfnamefont {S.~R.}\ \bibnamefont {Plissard}}, \bibinfo {author}
  {\bibfnamefont {E.~P. A.~M.}\ \bibnamefont {Bakkers}}, \ and\ \bibinfo
  {author} {\bibfnamefont {L.~P.}\ \bibnamefont {Kouwenhoven}},\ }\href
  {\doibase 10.1126/science.1222360} {\bibfield  {journal} {\bibinfo  {journal}
  {Science}\ }\textbf {\bibinfo {volume} {336}},\ \bibinfo {pages} {1003}
  (\bibinfo {year} {2012})}\BibitemShut {NoStop}%
\bibitem [{\citenamefont {Das}\ \emph {et~al.}(2012)\citenamefont {Das},
  \citenamefont {Ronen}, \citenamefont {Most}, \citenamefont {Oreg},
  \citenamefont {Heiblum},\ and\ \citenamefont {Shtrikman}}]{Das2012}%
  \BibitemOpen
  \bibfield  {author} {\bibinfo {author} {\bibfnamefont {A.}~\bibnamefont
  {Das}}, \bibinfo {author} {\bibfnamefont {Y.}~\bibnamefont {Ronen}}, \bibinfo
  {author} {\bibfnamefont {Y.}~\bibnamefont {Most}}, \bibinfo {author}
  {\bibfnamefont {Y.}~\bibnamefont {Oreg}}, \bibinfo {author} {\bibfnamefont
  {M.}~\bibnamefont {Heiblum}}, \ and\ \bibinfo {author} {\bibfnamefont
  {H.}~\bibnamefont {Shtrikman}},\ }\href {\doibase 10.1038/nphys2479}
  {\bibfield  {journal} {\bibinfo  {journal} {Nat. Phys.}\ }\textbf {\bibinfo
  {volume} {8}},\ \bibinfo {pages} {887Ð895} (\bibinfo {year}
  {2012})}\BibitemShut {NoStop}%
\bibitem [{\citenamefont {Sato}\ and\ \citenamefont
  {Fujimoto}(2009)}]{Sato2009}%
  \BibitemOpen
  \bibfield  {author} {\bibinfo {author} {\bibfnamefont {M.}~\bibnamefont
  {Sato}}\ and\ \bibinfo {author} {\bibfnamefont {S.}~\bibnamefont
  {Fujimoto}},\ }\href {\doibase 10.1103/PhysRevB.79.094504} {\bibfield
  {journal} {\bibinfo  {journal} {Phys. Rev. B}\ }\textbf {\bibinfo {volume}
  {79}},\ \bibinfo {pages} {094504} (\bibinfo {year} {2009})}\BibitemShut
  {NoStop}%
\bibitem [{\citenamefont {Fu}\ and\ \citenamefont {Kane}(2008)}]{Fu2008}%
  \BibitemOpen
  \bibfield  {author} {\bibinfo {author} {\bibfnamefont {L.}~\bibnamefont
  {Fu}}\ and\ \bibinfo {author} {\bibfnamefont {C.~L.}\ \bibnamefont {Kane}},\
  }\href {\doibase 10.1103/PhysRevLett.100.096407} {\bibfield  {journal}
  {\bibinfo  {journal} {Phys. Rev. Lett.}\ }\textbf {\bibinfo {volume} {100}},\
  \bibinfo {pages} {096407} (\bibinfo {year} {2008})}\BibitemShut {NoStop}%
\bibitem [{Note1()}]{Note1}%
  \BibitemOpen
  \bibinfo {note} {Conversely, it has been shown \cite {Sato2009,Qi2010} that
  the $\phi = \pi $ gives us the time-reversal invariant topological
  superconductor in the $h \to 0$ limit.}\BibitemShut {Stop}%
\bibitem [{\citenamefont {Read}\ and\ \citenamefont {Green}(2000)}]{Read2000}%
  \BibitemOpen
  \bibfield  {author} {\bibinfo {author} {\bibfnamefont {N.}~\bibnamefont
  {Read}}\ and\ \bibinfo {author} {\bibfnamefont {D.}~\bibnamefont {Green}},\
  }\href {\doibase 10.1103/PhysRevB.61.10267} {\bibfield  {journal} {\bibinfo
  {journal} {Phys. Rev. B}\ }\textbf {\bibinfo {volume} {61}},\ \bibinfo
  {pages} {10267} (\bibinfo {year} {2000})}\BibitemShut {NoStop}%
\bibitem [{\citenamefont {Schrieffer}(1983)}]{Schrieffer1983}%
  \BibitemOpen
  \bibfield  {author} {\bibinfo {author} {\bibfnamefont {J.}~\bibnamefont
  {Schrieffer}},\ }\href@noop {} {\emph {\bibinfo {title} {Theory of
  superconductivity}}},\ Advanced Book Program Series\ (\bibinfo  {publisher}
  {Advanced Book Program, Perseus Books},\ \bibinfo {year} {1983})\BibitemShut
  {NoStop}%
\bibitem [{Note2()}]{Note2}%
  \BibitemOpen
  \bibinfo {note} {Hence we predict that the fluctuation of $\phi $ will not
  contribute to the Hall conductivity.}\BibitemShut {Stop}%
\bibitem [{\citenamefont {Niu}\ \emph {et~al.}(1985)\citenamefont {Niu},
  \citenamefont {Thouless},\ and\ \citenamefont {Wu}}]{Niu1985}%
  \BibitemOpen
  \bibfield  {author} {\bibinfo {author} {\bibfnamefont {Q.}~\bibnamefont
  {Niu}}, \bibinfo {author} {\bibfnamefont {D.~J.}\ \bibnamefont {Thouless}}, \
  and\ \bibinfo {author} {\bibfnamefont {Y.-S.}\ \bibnamefont {Wu}},\ }\href
  {\doibase 10.1103/PhysRevB.31.3372} {\bibfield  {journal} {\bibinfo
  {journal} {Phys. Rev. B}\ }\textbf {\bibinfo {volume} {31}},\ \bibinfo
  {pages} {3372} (\bibinfo {year} {1985})}\BibitemShut {NoStop}%
\bibitem [{\citenamefont {Qi}\ \emph {et~al.}(2010)\citenamefont {Qi},
  \citenamefont {Hughes},\ and\ \citenamefont {Zhang}}]{Qi2010}%
  \BibitemOpen
  \bibfield  {author} {\bibinfo {author} {\bibfnamefont {X.-L.}\ \bibnamefont
  {Qi}}, \bibinfo {author} {\bibfnamefont {T.~L.}\ \bibnamefont {Hughes}}, \
  and\ \bibinfo {author} {\bibfnamefont {S.-C.}\ \bibnamefont {Zhang}},\ }\href
  {\doibase 10.1103/PhysRevB.81.134508} {\bibfield  {journal} {\bibinfo
  {journal} {Phys. Rev. B}\ }\textbf {\bibinfo {volume} {81}},\ \bibinfo
  {pages} {134508} (\bibinfo {year} {2010})}\BibitemShut {NoStop}%
\end{thebibliography}%

\end{document}